\documentclass{aa}
\usepackage[final]{changes}

\usepackage{graphicx}
\usepackage{hyperref}
\hypersetup{
    colorlinks=true,
    citecolor=blue,
    linkcolor=blue,
    filecolor=magenta,      
    urlcolor=blue,
}

\usepackage{txfonts}

\usepackage{natbib}
\bibpunct{(}{)}{;}{a}{}{,}

\def\kms  {km\,s$^{-1}$}

\def\bco {\ifmmode{^{12}{\rm CO}(J=2\to1)}\else{$^{12}{\rm
CO}(J=2\to1)$}\fi}

\def\hii  {{H\sc {\scriptsize II}}}
\def\mh     {H$_{2}$}

\def\Lsun{L$_{\odot}$}

\def\deg {$^{\circ}$}

\def\methanimine {CH$_2$NH}
\def\methanol {CH$_3$OH}

\def\water {H$_2$O}
\def\form   {H$_2$CO}
\begin{document}

\title{Discovery of methanimine (\methanimine) megamasers toward compact obscured galaxy nuclei}


   \author{M. D. Gorski
          \inst{1}
          \and
          S. Aalto\inst{1}
          \and
          J. Mangum\inst{2}
          \and
          E. Momjian\inst{3}
          \and
          J. H. Black\inst{1}
          \and
          N. Falstad\inst{1}
          \and
          B. Gullberg\inst{1}
          \and
          S. K{\"o}nig\inst{1}
          \and
          K. Onishi\inst{1}
          \and
          M. Sato\inst{1}
          \and
          F. Stanley\inst{4}
          }

   \institute{Department of Space, Earth and Environment, 
            Chalmers University of Technology, Onsala Space Observatory, 
            439 92 Onsala, Sweden \\
            \email{mark.gorski@chalmers.se}
        \and
            National Radio Astronomy Observatory, 
            520 Edgemont Road, 
            Charlottesville, VA 22903-2475, USA
        \and
            National Radio Astronomy Observatory, 
            P.O. Box O, 
            Socorro, NM 87801, USA
        \and
            Sorbonne Universit\'e, 
            UPMC Universit\'e Paris 6 \& CNRS, 
            UMR 7095, 
            Institut d'Astrophysique de Paris, 98b boulevard Arago,
75014 Paris, France
             }

   \date{recived \today }


\abstract{
We present the first search for the 5.29~GHz methanimine~(\methanimine)~$1_{10}-1_{11}$  transition toward a sample of galaxy nuclei.
We target seven galaxies that host compact obscured nuclei (CONs) with the  Karl G. Jansky Very Large Array. 
These galaxies are characterized by Compton-thick cores. 
\methanimine\ emission is detected toward six CONs.
The brightness temperatures measured toward Arp~220 indicate maser emission. 
Isotropic luminosities of the \methanimine\ transition, from all sources where it is detected, exceed 1~\Lsun\ and thus may be considered megamasers.
We also detect formaldehyde~(\form) emission toward three CONs.
The isotropic \methanimine\ luminosities are weakly correlated with the infrared luminosity of the host galaxy and strongly correlated with OH megamaser luminosities from the same galaxies.
Non-local thermodynamic equilibrium radiative transfer models suggest that the maser is pumped by the intense millimeter-to-submillimeter 
Our study suggests that  \methanimine\ megamasers are linked to the nuclear processes within 100~pc of the Compton-thick nucleus within CONs. 
}

  
\maketitle

\section{Introduction}

Interstellar methanimine (\methanimine) was first detected toward Sagittarius B2 (Sgr-B2) and is the simplest molecule to contain the carbon nitrogen double bond \citep{Godfrey1973}.
{The first  \methanimine\ maser emission was later discovered in Sgr-B2 by \citet{Faure2018}.}  
Methanimine is  the simplest of the ``imines,'' which are precursors to amino acids \citep{Danger2011}, and thus an important tracer of prebiotic chemistry in the universe.
We report the discovery of the first \methanimine\ megamaser and provide evidence for more \methanimine\ masers toward compact obscured nuclei (CONs).

\methanimine\ has been detected in both extragalactic and Galactic environments.
The first extragalactic detection of \methanimine\ was toward Arp~220  with the Arecibo radio telescope \citep{Salter2008}.
The line was observed in emission and hypothesized to be a maser.
A maser is a source of stimulated emission. 
It was also identified by \citet{Martin2006} in NGC~253 and in absorption toward PKS~1830-211 \citep{Muller2011}. 
In the Milky~Way, \methanimine\ has been detected toward the Galactic center \citep{Godfrey1973,Turner1991} with abundance enhancements toward high-mass star forming regions \citep{Dickens1997}.

Compact obscured nuclei are galaxies that host dusty and optically thick galactic centers \citep[e.g.,][]{Sakamoto2010,Gonzalez-Alfonso2012,Aalto2015b, Falstad2019, Falstad2021}.
Radiation at X-ray to millimeter wavelengths is strongly attenuated in these regions, where molecular gas column densities exceed $\rm{N(H_2)}=10^{25}$~cm$^{-2}$ \citep[e.g.,][]{Treister2010,Roche2015,Aalto2015a,Aalto2019} and dust temperatures are  $\gtrsim100$~K \citep[e.g.,][]{Sakamoto2013,Aalto2015a,Aalto2015b,Aalto2019}.
The frequency range  from 5 to 80 GHz is the only frequency range where these nuclei may be optically thin \citep{Barcos-Munoz2015,Sakamoto2017,Barcos-Munoz2018,Aalto2019}.
However, \citet{Martin2016} show that at the high frequency end of this spectral range, dust emission may still be opaque. 
For this reason, it is not known if an active galactic nucleus (AGN) or a nuclear starburst  powers the extreme infrared luminosities ($L_{\rm{IR}}~>10^{11}$~\Lsun) and outflows of these galaxies.

This paper presents a search for the 5.29~GHz \methanimine\ transition toward CONs.
\citet{Aalto2015b} found \methanimine\ emission at millimeter wavelengths ($\sim263$~GHz) toward two CONs: IC~860 and Zw~049.057.
They suggest that \methanimine\ may be an important tracer of the CON environment. 
Astrophysical masers are important probes of galaxy nuclei: for example  OH~\citep[e.g.,][]{Henkel1990}, \form\ \citep[e.g.,][]{Baan2017}, \water\ \citep[e.g.,][]{Herrnstein1999, Reid2009}, and \methanol\ \citep[e.g.,][]{Chen2015}.
Studies of these masers reveal outflows \citep[e.g., OH;][]{Baan1989}, molecular tori \citep[e.g.,][]{Lonsdale1998}, accretion disks of AGN \citep[e.g.,][]{Reid2009}, intense star formation \citep[e.g.,][]{Hagiwara2001,Brunthaler2009,Gorski2019}, and cloud-scale shocks \citep[e.g.,][]{Ellingsen2017, Gorski2017,Gorski2018}.
Because masers often trace specific conditions within the interstellar medium, finding new  maser species may be critical to unveiling what is hidden behind the thick dust veils of CONs.

We report the detection of the 5.29~GHz \methanimine~$1_{10}-1_{11}$ transition toward six galaxies: Arp~220, IC~860, Zw~049.057, IRAS~17208$-$0014, IRAS~17578$-$0400, and NGC~4418.
Toward Arp~220 we provide clear evidence for a \methanimine\ megamaser. 

\begin{table*}
\centering
\caption{Observational parameters.}
    \begin{tabular}{llllllll}
    \hline\hline
    Project Code & Source &RA & Dec. &$z$ & Complex Gain Calibrator & Channel Width \\
    &   &(J2000)   &(J2000)   &   &   &(kHz)  \\
    \hline  
    20A-501 & Zw~049.057 & 15:13:13.10 & +07:13:32.0 & 0.0130        & J1504+1029 & 125 \\
    20A-501 & IRAS~17208$-$0014 & 17:23:21.95 & $-$00:17:00.9 & 0.0428   & J1743$-$0350 & 125 \\
    20A-501 & IRAS~17578$-$0400 & 18:00:31.85 & +23:30:10.5 & 0.0134   & J1743$-$0350 & 125 \\
    20A-501 & NGC~4418  & 12:26:54.62 & $-$00:52:39.4 & 0.0073         & J1224+0330 & 125 \\
    20A-501 & IRAS~22491$-$1808 & 22:51:49.31 & $-$17:52:24.0 & 0.0778   & J2246$-$1206 & 125 \\
    15A-398 & IC~860 & 13:15:03.53 & +24:37:07.9 & 0.0129            & J1504+1029 & 125 \\
    11A-231 & Arp~220 & 15:34:57.27 & +23:30:10.5 & 0.0181           & J1513+2338 & 250 \\

    \hline\hline
    \end{tabular}
    \label{tab:obsprop}
    \vspace{-6pt}

\end{table*}

\begin{table*}
\centering
\caption{Data cube parameters.}
    \begin{tabular}{lllllll}
    \hline\hline
     Source & Weighting & Molecule & Channel Width  & RMS noise & Beam Dimensions & Position Angle\\
    & & & \kms  & mJy~beam$^{-1}$   &  &\\
    \hline 
    Zw 049.057 & 0.5      & \methanimine & 20      & 0.37 & 5\farcs27$\times$3\farcs97 & \phantom{$-$}49.94\deg\\
                   & & \form        & 20      & 0.38 & 5\farcs91$\times$4\farcs19 & \phantom{$-$}51.6\deg\\
    IRAS~17208$-$0014 & 0.5 & \methanimine & 100      & 0.25 & 5\farcs04$\times$4\farcs81 & \phantom{$-$}15.2\deg\\
                   & & \form        & 100      & 0.24 & 4\farcs87$\times$4\farcs14 & \phantom{$-$}13.4\deg\\
    IRAS~17578$-$0400 & 0.5 & \methanimine & 20      & 0.37 & 4\farcs82$\times$3\farcs56 & \phantom{$-$0}6.5\deg\\
                   & & \form        & 50      & 0.31 & 5\farcs20$\times$3\farcs98 & \phantom{$-$0}7.6\deg\\
    NGC~4418 & 0.5        & \methanimine & 50      & 0.48 & 2\farcs06$\times$1\farcs18 & \phantom{$-$}48.4\deg\\
                   & & \form        & 50      & 0.49 & 2\farcs32$\times$1\farcs28 & \phantom{$-$}48.0\deg\\
    IRAS~22491$-$1808 & 0.5 & \methanimine & 50      & 0.44 & 2\farcs24$\times$1\farcs19 & \phantom{$-$}21.5\deg\\
                    & & \form       & 50      & 0.44 & 2\farcs72$\times$1\farcs38 & \phantom{$-$}22.4\deg\\
    IC~860 & 0.5         & \methanimine & 66      & 0.23 & 0\farcs42$\times$0\farcs35 & $-$63.4\deg\\
                   & & \form        & 66      & 0.22 & 0\farcs43$\times$0\farcs38 & $-$63.4\deg\\
    Arp~220 & 0.5        & \methanimine & 18      & 0.19 & 0\farcs36$\times$0\farcs28 & $-$67.7\deg\\
    Arp~220 & uniform & \methanimine & 18      & 0.33 & 0\farcs32$\times$0\farcs26 & $-$70.6\deg\\
    \hline\hline
    \end{tabular}
    \label{tab:cubeprop}
    \vspace{-6pt}
\end{table*}

\section{Observations}

Zw~049.057, IRAS~17208$-$0014, and IRAS~17578$-$0400 were observed with the Karl G. Jansky Very Large Array (VLA) in C configuration, and NGC~4418 and IRAS~22491$-$1808 were observed in B configuration (Project Code 20A-501).
Individual 128~MHz wide subbands were placed to target the 4.8~GHz \form, 5.29~GHz \methanimine, and 6.7~GHz \methanol\ masers with 125~kHz wide channels.
Arp 220 and IC 860 were both observed with the VLA in A configuration (project codes 11A-231 and 15A-398), with 250 kHz and 125 kHz wide channels, respectively.

The data were calibrated and imaged in the Common
Astronomy Software Applications (CASA) package version
5.0.0 (McMullin et al. 2007).
With two exceptions, 3C286 (flux density = 7.47 Jy at 5.1 GHz) was observed as the bandpass and flux density scale calibrator.
For Arp~220, J1602+3326 was observed as the bandpass calibrator, and for IRAS~22491$-$1808 3C147 (flux density = 6.74 Jy at 5.5 GHz) was observed as the bandpass and flux density scale calibrator.
The details of the observations are summarized in Table \ref{tab:obsprop}.

The final data cubes were produced using a Briggs robustness value of 0.5 in the CASA task tclean, except for Arp~220. 
For Arp~220 two data cubes were made, one with a Briggs robust value of 0.5 and one uniformly weighted for the smallest synthesized beam.
The frequency axes were resampled to a velocity resolution of 18~\kms to 100~\kms and RMS values ranged from 0.19~mJy~beam$^{-1}$ to 0.49~mJy~beam$^{-1}$.
All velocities are reported in the kinematic local standard of rest (LSRK) frame. 
Continuum subtraction was performed  by selecting line-free channels and fitting with a polynomial of order 1 in the image domain. 
The continuum flux density at the line peak is determined by fitting a polynomial of order 1 to the entire 128 MHz subband.  5\% of channels at the band edges, and suspected line containing channels, were ignored.
The properties of each image cube are listed in Table \ref{tab:cubeprop}.

\section{Results}
\begin{figure*}
    \centering
    \includegraphics[width=0.95\textwidth]{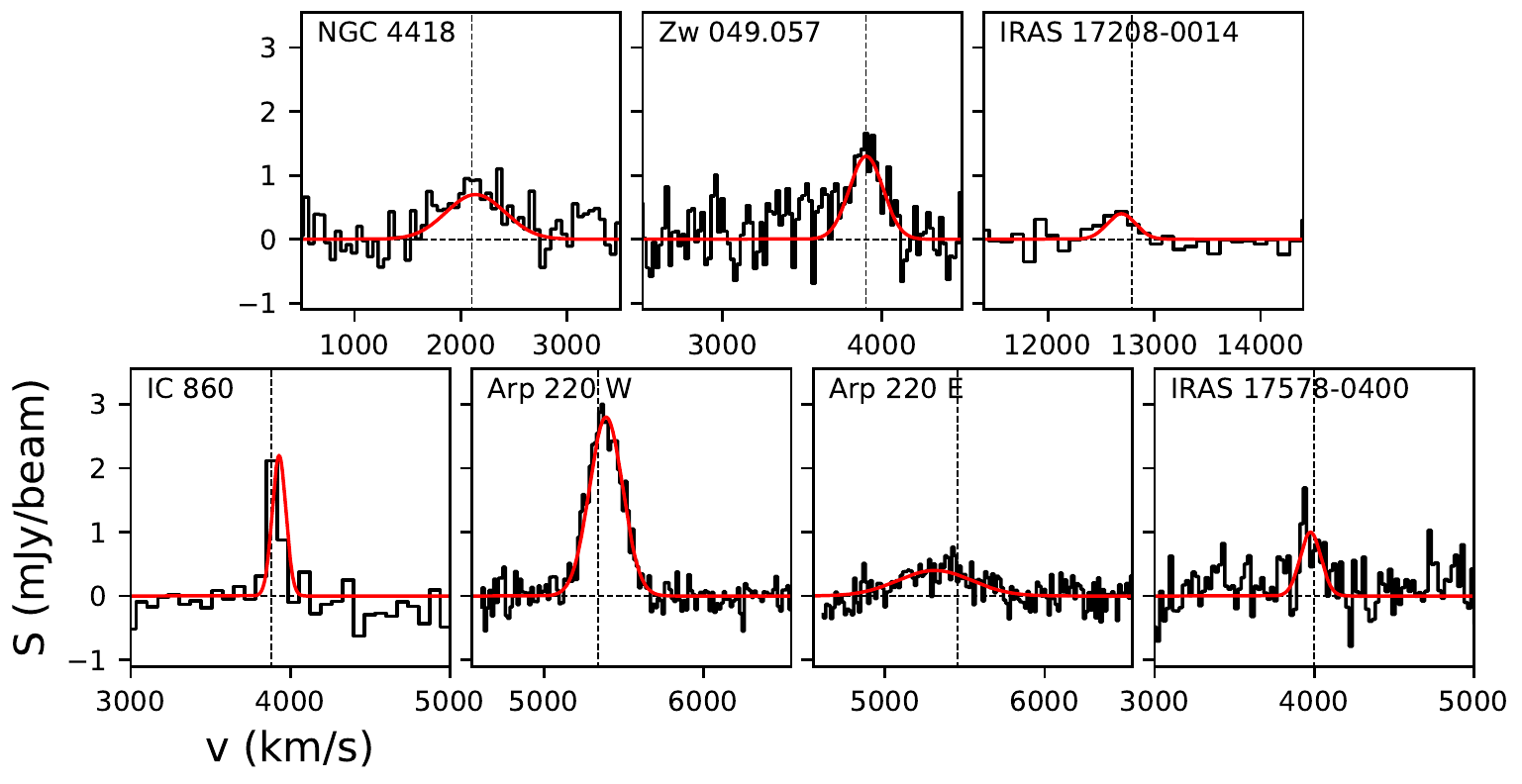}
    \caption{Observed \methanimine\ spectra toward six CONs, including both nuclei in Arp~220. 
    The horizontal black dashed line shows zero flux,  the vertical black dashed line shows the systemic velocity of each galaxy, and the red line indicates the Gaussian best fit. Systemic velocities are adopted from \citet[][]{Aalto2015a} for IC~860 and Zw~049.057, from \citet{Martin2016} for Arp~220, from \citet{Sakamoto2013} for NGC~4418, from \citet{GarciaBurillo2015} for IRAS~17208$-$0014, and from \citet{Falstad2021} for IRAS~17578$-$0400.
    }
    \label{fig:ch2nhlines}
\end{figure*}

\begin{figure*}
    \centering
    \includegraphics[width=0.95\textwidth]{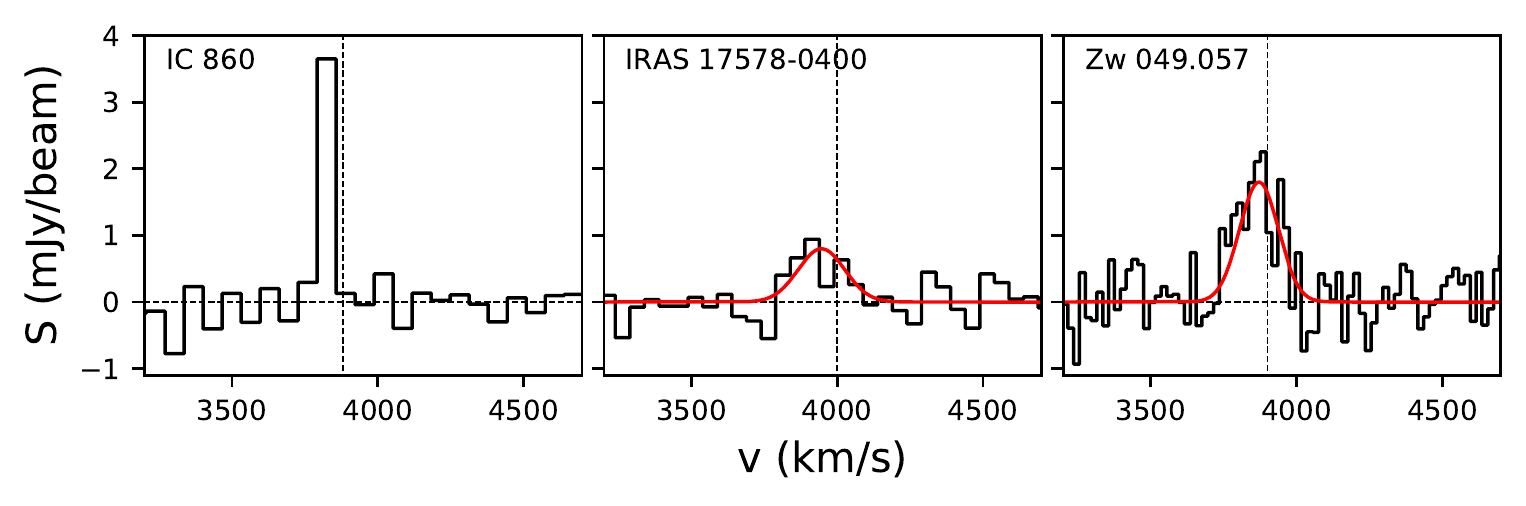}
    \caption{Observed \form\ spectra toward three CONs.
    The horizontal black dashed line shows zero flux,  the vertical black dashed line shows the systemic velocity of each galaxy, and the red line indicates the Gaussian best fit. Systemic velocities are adopted from \citet[][]{Aalto2015a} for IC~860 and Zw~049.057 and from \citet{Falstad2021} for IRAS~17578$-$0400.}
    \label{fig:h2colines}
\end{figure*}

 The 5.29 GHz \methanimine~$1_{10}-1_{11}$ transition is detected in emission toward the nuclei of six galaxies:  Arp~220, IC~860, Zw~049.057, IRAS~17208$-$0014, IRAS~17578$-$0400, and NGC~4418. 
 In addition, we report a new detection of the 4.8~GHz formaldehyde (\form) transition toward IRAS~17578$-$0400.
 %
 This increases the number of galaxies with known \form\ 4.8~GHz emission from five (UGC~5101, IC~860, Arp~220, NGC 3079, and Zw~049.057 \citealp{Mangum2008,Mangum2013}) to six.
 In all observations the \methanimine\ and \form\ lines are spatially unresolved. 
 For each galaxy the \methanimine\ and \form\ transitions were fit with a Gaussian profile. 
 Table \ref{tab:lineprop} presents the derived peak flux density ($\rm{S_p}$), continuum flux density at the line peak, velocity full width at half maximum (FWHM),  center velocity, integrated flux density, and peak brightness temperature ($\rm{T_{pk}}$)  of the \methanimine\ and the 4.8 GHz \form{} transition. 
 Figure \ref{fig:ch2nhlines} shows the observed line profiles.
 For all galaxies the FWHM of the line is $>100$\,\kms{}  except IC~860 where both the \form\ and \methanimine\ transitions are spectrally unresolved (FWHM$<66$\,\kms).
 If we assume isotropic radiation and that the line width is nonrelativistic:
 \begin{equation}
     \begin{aligned}
     L &=4 \pi D^2 \int S d \rm{\nu} \\
     \rm{d}\nu &= \frac{\rm{dv}}{c}\nu_{0}
     \end{aligned}
 ,\end{equation}
 where $\nu_{0}$ is the rest frequency of the line, D is the luminosity distance to the object, and the integral is  the integrated line flux density.
 The observed luminosity of an emission line is thus:
\begin{equation}
     \begin{aligned}
     L &=4 \pi D^2 \frac{\nu_0}{c}  \int S d\rm{v} \\
     \end{aligned}
 .\end{equation}
 The luminosity of the \methanimine\ transition, in units common to observational extragalactic astronomy, is calculated as
 \begin{equation}
     L_{\rm{CH_2NH}}[L_\odot]=5.53\times10^{-3}\times (D[\rm{Mpc}])^2\times \int S d\rm{v}[\rm{Jy\,km\,s^{-1}}]
 \end{equation}
 and for formaldehyde as \begin{equation}
     L_{\rm{H_2CO}}[L_\odot]=5.04\times10^{-3}\times (D[\rm{Mpc}])^2\times \int S d\rm{v}[\rm{Jy\,km\,s^{-1}}]
 .\end{equation}
 The integral represents the integrated line flux density in units of Jy\,\kms, and $D$ is the distance in megaparsecs. 
 Distances are adopted from \citet{Sanders2003}. 
 We provide upper limits toward the  galaxies where neither \methanimine\ nor \form\ was detected.
 Among the detected emission, \methanimine\ luminosity varies between 2.9~\Lsun\ and 27~\Lsun\, and the \form\ luminosity varies between 2.8~\Lsun\ and 5.3~\Lsun. 
 
\begin{table*}
\caption{Emission line parameters.} 
    \begin{tabular}{lrrrrrrr}
    \hline\hline
     Source & $S_{\rm cont}$   & $S_{\rm p}$ & $\Delta v_{\rm FWHM}$ & $v_{\rm Center}$ & $\int{S}\,d\nu$ & $T_{\rm pk}^a$& Luminosity\\
               & [mJy] & [mJy]     & [\kms{}]& [\kms{}] & [Jy \kms{}]& [K] &[\Lsun{}]\\
    \hline 
    \textbf{\methanimine} \\
    \hline
    Zw 049.057 & 29.55$\pm$0.06    & 1.25$\pm$0.19 & 238$\pm$46 & 3901$\pm$19 & 0.31$\pm$0.05 & \phantom{0}2.61$\pm$0.40 & \phantom{0}6.22$\pm$0.99 \\
    
    IRAS 17208$-$0014 & 55.14$\pm$0.18    &  \phantom{0}0.44$\pm$0.13 & 375$\pm$139 & 12690$\pm$56  & 0.14$\pm$0.05 & \phantom{0}0.81$\pm$0.18      & 32.81$\pm$10.01  \\
    
    IRAS 17578$-$0400 &31.85$\pm$0.05    & \phantom{0}0.90$\pm$0.21  & 158$\pm$53  & 3977$\pm$19 & 0.15$\pm$0.03 & \phantom{0}2.29$\pm$0.53  & \phantom{0}3.04$\pm$0.70 \\
   
    NGC~4418 & 22.94$\pm$0.08   & 0.78$\pm$0.13 & 689$\pm$146 & 2138$\pm$56 & 0.57$\pm$0.10 & 14.02$\pm$2.34 & \phantom{0}3.50$\pm$0.60 \\
    
    IRAS 22491$-$1808 & \phantom{0}3.22$\pm$0.04    &<1.32          & -           & -            & <0.066          & <19.7          & <39.30         \\
    
    IC 860& 22.24$\pm$0.03   & \phantom{0}2.2$\pm$0.23  & <68         & 3980$\pm$68  & 0.15$\pm$0.02          & 657$\pm$69     & \phantom{0}2.88$\pm$0.38 \\
    
    Arp 220 W &84.00$\pm$0.19    & 2.81$\pm$0.07 & 254$\pm$8\phantom{0}   & 5390$\pm$28  & 0.74$\pm$0.02 & 1216$\pm$28   & 26.80$\pm$0.65 \\
    
    Arp 220 E& \phantom{0}54.6$\pm$0.19    & 0.46$\pm$0.05 & 535$\pm$70  & 5348$\pm$24  & 0.26$\pm$0.03 & \phantom{0}229$\pm$22     & \phantom{0}9.47$\pm$0.99  \\
    \hline 
    \textbf{\form{}} \\
    \hline
    Zw 049.057& 31.09$\pm$0.09    & 1.79$\pm$0.22  & 164$\pm$29 & 3921$\pm$12  & 0.31$\pm$0.04 & \phantom{0}3.74$\pm$0.47 & \phantom{0}6.28$\pm$0.94 \\
    
    IRAS 17208$-$0014& 58.01$\pm$0.16    &<1.23          & -          & -            & <0.031          & <3.19         & <5.19         \\
    
    IRAS 17578$-$0400& 34.10$\pm$0.10   & 0.77$\pm$0.18  & 177$\pm$77 & 3998$\pm$32 & 0.15$\pm$0.04 & \phantom{0}1.95$\pm$0.46 & \phantom{0}5.45$\pm$1.98 \\
   
    NGC~4418& 24.82$\pm$0.08    & <1.47          & -          & -            & <0.074          & <26.07        & <0.41         \\
    
    IRAS 22491$-$1808& \phantom{0}3.40$\pm$0.02   & <1.32          & -          & -            & <0.066          & <18.5         & <35.81        \\
    
IC 860& 22.79$\pm$0.16   & \phantom{0}3.7$\pm$0.20    & <66        & 3852$\pm$66  & 0.24$\pm$0.02          & 1191$\pm$71   & \phantom{0}4.29$\pm$0.35         \\
    \hline\hline
    \end{tabular}
    \vspace{-6pt}
    \tablefoot {Uncertainties reported in this table are determined from  the covariant matrix of the Gaussian fit.}
    \tablefoottext{a}{The measured brightness temperatures are all lower limits as the transitions are unresolved toward all galaxies.}
    \label{tab:lineprop}
\end{table*} 

\section{Discussion}
\subsection{Evidence for maser emission}

Toward all galaxies the \methanimine\ and  \form\ transitions are spatially unresolved, and thus the brightness temperatures reported in this paper are lower limits.
Even galaxies observed in the most extended configuration of the array, A configuration with synthesized beam dimensions 0\farcs36$\times$0\farcs28\ (Table~\ref{tab:cubeprop}), are unresolved. 
In the case of Arp~220, the measured brightness temperature of \methanimine\ toward the western core is $>1216\pm28$~K in the Briggs weighted image cube.
By uniformly weighting the visibilities the spatial resolution of the \methanimine\ image cube is improved to 0\farcs32$\times$0\farcs26 and the peak brightness temperature is measured to be 1470$\pm$68~K (3.2$\pm$0.3~mJy).
Attempting to de-convolve the \methanimine\ source associated with the western nucleus of Arp~220 with the CASA task {\sc{imfit}}, we found an unresolved source. The line emission in Arp 220 must be superthermal because its peak brightness temperature greatly exceeds the physical temperatures (gas kinetic and dust), $\leq 300$ K, derived for Arp 220 and other CONs \citep{Sakamoto2010,Aalto2019,Zschaechner2016}.

Furthermore, the physical temperatures toward CONs do not appear to exceed 300~K, for example in NGC~4418 \citep{Sakamoto2010}, IC~860 \citep{Aalto2019}, and Arp~220 \citep{Zschaechner2016}. 
High brightness temperatures measured toward IC~860 and Arp~220 can be attributed to maser emission; however, the remaining detections require higher angular resolution observations to confirm their masing nature.
If the \methanimine\ emission traces the innermost region (<50~pc) of the CONs, the brightness temperatures may be significantly higher.
\citet{Aalto2015b,Aalto2019} show that the lines of vibrationally excited HCN (HCN-VIB) toward CONs trace structures smaller than 0\farcs2 ($\lesssim$60pc).
\citet{Costagliola2013} show that the CON of NGC~4418 has an angular diameter less than 0\farcs3 (50~pc). 
Assuming the CONs are smaller than 50~pc and the \methanimine\ masers probe the CONs, then the brightness temperatures toward NGC~4418, Zw~49.057, IRAS~17208$-$0014, IC~860, and Arp~220 W would respectively be 380~K, 2000~K, 6700~K, 3400~K, and 7900~K.
Thus, it is likely that the 5.29~GHz line is masing toward all the CONs we have observed.

The measured isotropic luminosity of the \methanimine\ line is $>2$~\Lsun\ and $\lesssim30$~\Lsun\ in all cases where it is detected. 
A megamaser is defined as 10$^6$ times as luminous as the average Milky Way maser of the same transition.
For example the cutoff for \water\ megamasers is 20~\Lsun\ (see \citealt{Hagiwara2001} for a description of this nomenclature).
The Milky Way detections of the 5.29 GHz \methanimine~$1_{10}-1_{11}$ transition  have luminosities of $\sim1.1\times10^{-6}$~\Lsun\ and $\sim0.5\times10^{-6}$~\Lsun\ \citep{Faure2018} if one adopts a distance of 8.3~kpc \citep{Reid2014}.
The least luminous detection in our study is toward IC~860, where the line luminosity is  2.88~\Lsun, and the most luminous is toward the western core of Arp220, where the line luminosity is  26.8~\Lsun. 
All the galaxies we have observed with detections of the \methanimine\ line qualify as megamasers assuming Sgr-B2 masers from \citet{Faure2018} represent the more luminous end of the distribution of Milky Way \methanimine\ masers.

\subsection{Non-LTE modeling}
\label{sec:LVG}
\cite{Baan2017}  
show that population inversions of H$_2$CO can be maintained by infrared  pumping based on calculations with the nonlocal thermodynamic equilibrium (non-LTE) radiative transfer code {\tt RADEX} \citep{vanderTak2007}.  
Such models of maser emission based on a simple escape probability approximation can be reliable until the maser saturates at optical depths $\tau \leq -1$. 
However, \citet{Baan2017} did not discuss whether the luminosity implied by their assumed blackbody continuum at $T_{\rm bb} = 50$ K was consistent with the observed infrared emission on the same small angular scale, nor did they include the near- and mid-infrared radiation that is known to be present in galaxies' luminous CONs, and that must also excite vibrational transitions in molecules such as H$_2$CO and CH$_2$NH.
\citet{Faure2018} recently investigated the CH$_2$NH maser emission in the Galactic-center molecular cloud complex Sgr B2. 
To do this, they computed rate coefficients for collisional excitation of  CH$_2$NH by para-H$_2$ for the 15 lowest rotational levels of methanimine at temperatures up to 30 K. 
However, their {\tt RADEX} analysis of the methanimine emission apparently omitted the infrared continuum radiation of the source and ignored radiative excitation processes except for the 2.7 K cosmic background radiation. 
\citet{Faure2018} did demonstrate a robust population inversion with excitation temperature $T_{\rm ex} = -0.48$ in the $1_{10} - 1_{11}$ transition at kinetic temperature $T=30$ K and density of para hydrogen $10^4$ cm$^{-3}$.

The physical conditions in CONs of galaxies may be more extreme than those explored in previous analyses of centimeter-wave masers in H$_2$CO and CH$_2$NH. 
Not only are the temperatures and densities higher than $30$ K and $10^4$ cm$^{-3}$, respectively, but the continuous radiation from centimeter wavelengths to the near-infrared filling these regions is extremely intense
(cf. 
\citealt{Sakamoto2013},  
\citealt{Aalto2015a},  
\citealt{Aalto2015b},  
\citealt{Gorski2018},     
\citealt{Mangum2019}).  
To explore the excitation of weak maser emission in such regions, we calculated non-LTE radiative transfer models with a code that fully incorporates all the features originally intended for {\tt RADEX} as described by \cite{vanderTak2007}.  
The new code, {\tt GROSBETA}, retains the simplified mean-escape probability treatment of radiative transfer but allows for arbitrary continuum spectral energy distributions, incorporates chemical formation-pumping where appropriate, and solves for many different molecules in the same computation. 
\citet{Tabone2021} describe the code and its application to superthermal OH emission.  
In order to illustrate the full range of radiative effects on the excitation of a molecule such as methanimine, we expanded the spectroscopic data files to include all fundamental vibrational transitions and explored a range of plausible infrared continua (see the appendices). 
The non-LTE computation solves for the steady-state density in each vibration-rotation level of a molecule subject to collisional
excitation at kinetic temperature $T_k$ and number densities of the main collision partners, for example $n({\rm H}_2)$. 
The other input parameters are the internal brightness of continuum radiation $J_{\nu}$ with dimensions [Jy sr$^{-1}$], the path length through the source $R$, and the molecular column density $N$ [cm$^{-2}$] over the FWHM of the line-of-sight velocity distribution, $\Delta V$ [km s$^{-1}$]. 
The effective solid angle of the source is $\Omega = \pi (R/D_{\rm L})^2$, where $D_{\rm L}$ is the luminosity distance. 
The observable flux in a line is given by
\begin{equation}
  f_{\nu} = \Omega \Bigl( I_{\nu,{\rm core}} \exp\bigl(-\tau_{\nu}\bigr) + B_{\nu}(T_{\rm ex}) \bigl(
  1 - \exp(-\tau_{\nu}) \bigr)\Bigr)
 \label{jb1}
,\end{equation}
where $I_{\nu,{\rm core}}$ is the surface brightness of the continuum in the core, $B_{\nu}(T_{\rm ex})$ is the Planck function evaluated at the excitation temperature of the line, and $\tau_{\nu}$ is the optical depth in the line. 
The first term represents the amplification (absorption) of the continuum radiation when the optical depth is negative (positive) and the second term is the self-emission of the molecular source. 
  
We take the formaldehyde and methanimine maser emission in IC 860 as a test case.
A crucial part of the non-LTE excitation calculation is the specification of the internal radiation field. 
For IC 860 we adopt observed fluxes as collected in the NASA/IPAC Extragalactic Database (NED)\footnote{https://ned.ipac.caltech.edu/, NED is funded by the US National Aeronautics and Space Administration and operated by the California Institute of Technology.}. 
The observed flux densities from meter to centimeter wavelengths are well-fitted by a flat power law spectrum, $S_{\nu} \propto \nu^{-0.215}$; therefore, we assume that the power-law component is very compact and contained entirely within the $0.42\times 0.35$ arcsecond projected beam area of the VLA observations presented here.
The corresponding solid angle is $\Omega = 3\times 10^{-12}$ sr. 
Most of the power in the observed spectrum of IC 860 is contained in the submillimeter and infrared  region (frequencies $3.5\times 10^{12}$ to $6.5\times 10^{13}$ Hz). The integrated flux density over
this frequency interval, 7.1 W m$^{-2}$, corresponds to a luminosity of $6.3\times 10^{11}$ L$_{\odot}$. 
The observed power is thought to be short-wavelength (visible, ultraviolet, X-ray) light from a central starburst and/or AGN that has been absorbed and reradiated by surrounding dust.  
Owing to the lack of far-infrared measurements at sub-arcsecond resolution, we do not know what fraction of the observed infrared power, $\varphi$, is contained within the solid angle $\Omega$ of the centimeter-wave radio source. 
However,  the molecules within $R=53$~pc of the center of IC 860 must be exposed to a mean brightness $J_{\nu}$ given by
\begin{equation}
  J_{\nu} \Omega  =  22.39 \Bigl({{\nu}\over{5.29 {\mathrm GHz}}}\Bigr)^{-0.215} 
10^{-0.4 A_{\lambda}} + \varphi  S_{\nu,{\rm obs}} \;\;\; {\mathrm mJy},
\label{jb2}
\end{equation} 
where $A_{\lambda}$ is the extinction in magnitudes for a standard interstellar extinction law, 
normalized to visual (550 nm wavelength) extinction $A_V=400$ mag. 
Including the extinction ensures that the power-law component does not exceed the observed central infrared radiation.
The model radiation field is presented in Appendix A.2. It can be useful to define the equivalent radiation brightness temperature $T_{\rm rad}$ such that the mean internal brightness is
\begin{equation}
J_{\nu} = B_{\nu}(T_{\rm rad}) \;\;\; ,     
\end{equation}
together with a function 
\begin{equation}
  y(\nu) = \Bigl( \exp(h\nu / kT_{\rm rad}) - 1\Bigr)^{-1} \;\;\;  .
\end{equation}
For any transition at frequency $\nu$ with spontaneous transition probability $A_{u,\ell}$ from upper state $u$ of statistical weight $g_u$ to lower state $\ell$ with statistical weight $g_{\ell}$, the rate of stimulated emission in this radiation field is $y(\nu) A_{u,\ell}$, while the rate of absorption is $y(\nu) g_u A_{u,\ell}\, /\, g_{\ell}$. 
For example, the CH$_2$NH transition $1_{10} - 1_{11}$ at 5.29 GHz has $A_{u,\ell} = 1.55\times 10^{-9}$ s$^{-1}$, while the value of 
$T_{\rm rad} = 8791$ K in the power-law component of the internal radiation field, so that $y=34625$ and the pumping rate $y A_{u,\ell} = 5.35\times 10^{-5}$ s$^{-1}$. 
Examples of pumping rates for two values of $\varphi = 1.0$ and $0.1$ are listed for a number of transitions in Table \ref{tab:pumping}.
This shows that induced absorption out of $J_{K_{\rm a}K_{\rm c}} = 1_{10}$, the upper level of the 5.29 GHz transition, is faster than spontaneous decay at 5.29 GHz for various rotational and vibration-rotation transitions at frequencies between 133 GHz and 50 THz. 
The highest pumping rates occur in the millimeter-to-submillimeter wavelength part of the spectrum. 
In the adopted radiation field, the vibration-rotation transitions in the infrared are not very important in the excitation of the $1_{10}$ and $1_{11}$ levels. 
Comparison of the rates also suggests why the centimeter-wave continuum enters the interpretation in two distinct ways:  (1) the internal 
radiation (Eq. \ref{jb2}) felt by the molecules is so strong that stimulated emission in the 5.29 GHz transition is much faster than spontaneous emission, and (2) the observable continuum emission is strong and must be taken into account in description of the line emission (Eq. \ref{jb1}). 
Finally, it is useful to compare collisional de-excitation rates with the induced radiative rates. The  $1_{10}$ level has a total spontaneous decay rate of $5.64\times 10^{-5}$ s$^{-1}$ mostly in the 166.85 GHz transition. 
The few collision rates presented by \citet{Faure2018} suggest downward collisional rate coefficients from $1_{10}$ on the order of $10^{-11}$ cm$^3$ s$^{-1}$; therefore, in the absence of any radiative couplings to higher states, collisions could dominate the excitation of $1_{10}$ at densities rather greater than $n({\rm H}_2) \sim 10^6$ cm$^{-3}$. 
With a realistic description of the radiation environment, on the other hand, we see that radiative pumping out of $1_{10}$ occurs with a rate on the order of $y A \sim 10^{-3}$~s$^{-1}$. 
Thus, collisions are unlikely to thermalize the populations of low-excitation levels such as $1_{10}$ unless the hydrogen density exceeds $10^8$ cm$^{-3}$, although collisions can suppress the population inversion at lower densities, on the order of $10^6$ cm$^{-3}$.

\begin{figure*}
    \centering
    \includegraphics{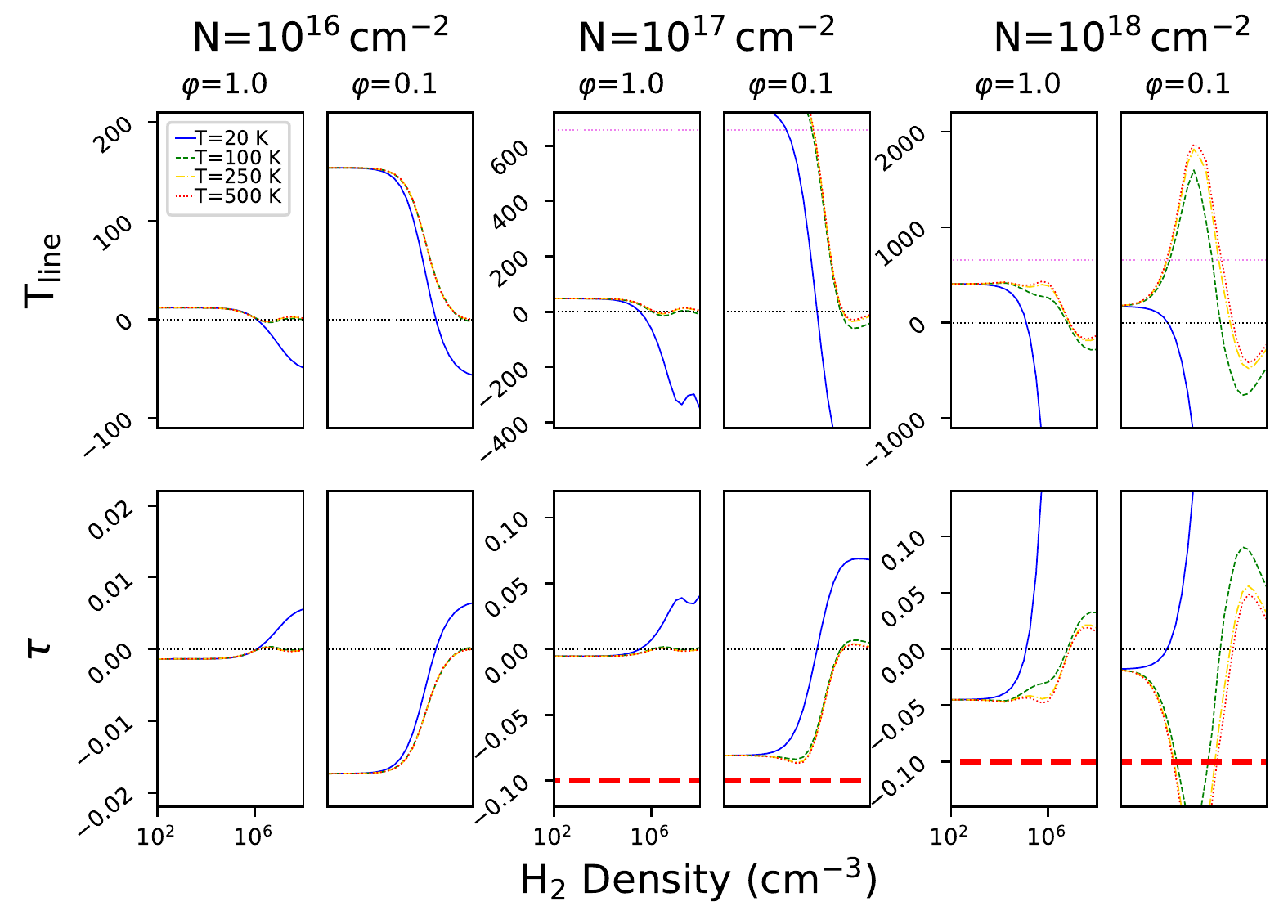}
    \caption{Non-LTE models showing radiative excitation of the 5.29~GHz \methanimine\ line. The top row of plots shows the line radiation temperature, and the bottom row shows the optical depth. The horizontal axis in all plots is the \mh\ density.  The horizontal black dotted line indicates a  vertical axis value of 0.0 on each plot. The violet horizontal dotted line indicates the observed brightness temperature toward IC~860. The horizontal thick red dashed line indicates the optical depth limit indicated by \citet{vanderTak2007} as line brightness temperatures may not be trusted if the optical depth is less than -0.1. Column densities for each pair of $\varphi$ are labeled at the top of each pair of columns. Temperatures of the molecular gas are labeled in the upper left corner. The line width input to the mean escape probability approximation is 100~\kms. The maser action of \methanimine\ in an infrared radiation field is observed in all cases from $10^{2}~\rm{cm}^{-2}$ to $\sim10^{5}~\rm{cm}^{-2}$.}
    \label{fig:radex_w_background}
\end{figure*}

The non-LTE treatment of excitation and radiative transfer includes all processes, such as those outlined in the preceding paragraph. 
Thus we can see what parameter space allows for a CH$_2$NH maser of the observed intensity in IC~860.  
The results of these models are shown in Fig. \ref{fig:radex_w_background}.
The models that reproduce the observed line brightness temperature, $T_{\rm line} \geq 657$ K, have total densities  in the range $n({\rm H}_2)=10^2$ to $10^5$ cm$^{-3}$. The solutions are insensitive to kinetic temperature, which reflects the dominance of radiative pumping in the excitation.
Large column densities and abundances are needed to reproduce the line intensity when $\varphi = 1$. 
At lower values (e.g., $\varphi \leq 0.1$), the required column density of CH$_2$NH is much lower, $\gtrsim1\times 10^{16}$ cm$^{-2}$, so that the required abundance varies inversely with the hydrogen density. 
Population inversion and weak amplification ($\tau \approx -0.1$) is maintained over a wide extent of parameter-space, producing line strengths of $\gtrsim100$~K,  without any need for delicate tuning of density or abundance. 

It should be possible to constrain the value of  $\varphi$  with further observations at millimeter-to-submillimeter wavelengths, both in lines and continuum, with sub-arcsecond angular resolution. 
For example, the model predicts that the $3_{03}-2_{12}$ transition at 35.055 GHz will also be a maser under the same conditions that explain the 5.29 GHz line intensity, but with $T_{\rm line} \sim 2000$ K, 100 K, and 25 K at $\varphi = 1, 0.1$, and $0.01$, respectively. 
Numerous CH$_2$NH transitions at frequencies 200 to 300 GHz are predicted to appear strongly in  emission at $\varphi = 1$, but to go into absorption with $\tau \sim 1$ when $\varphi \leq 0.1$. 
In particular, the predicted flux ratio of the $7_{16} - 7_{07}$ transition (frequency $250.162$ GHz, excitation energy $E_u/k=97$ K) and the $4_{04} - 3_{03}$ transition (at $254.685$ GHz with $E_u/k=31$ K) is $S_{250}/S_{254}\approx 2.5$ when $\varphi=1$, but becomes negative when the radiation scaling falls to $\varphi = 0.1$. 
These numbers refer to fluxes measured in the same effective solid angle as the centimeter-wave masers, $\Omega = 3\times 10^{-12}$.

In summary, panchromatic, non-LTE radiative transfer models of both CH$_2$NH reproduce the 
observed fluxes of centimeter-wave maser emission at {0}\farcs{4} angular resolution in IC~860. 
The population inversions and amplification are robust over a range of densities and temperatures. 
The compact, intense centimeter-wave continuum emission plays a major role both in the excitation of the masering levels and as the background flux that is amplified in the lines.
The molecular excitation may be dominated by radiative processes over the entire parameter space that sustains the required population inversions in both CH$_2$NH and H$_2$CO. 
The models predict fluxes of additional strong lines at
mm wavelengths, which could be used to constrain better the density and abundances.

\subsection{Comparing \methanimine\ masers in CONs with other maser species }

\subsubsection{\water\ masers}

The seven galaxies reported in this paper are all classified as CONs, and \methanimine~$1_{10}-1_{11}$ emission is detected toward six of these galaxies.     
Surveys for megamasers in AGN have low success rates, such as \citet{Sato2005}, where 90 Seyfert 2 or LINER galaxies were surveyed resulting in a single \water\ megamaser detection.
Of the $>2800$ galaxies surveyed for \water\ megamasers, 178 have clear detections \citep{Braatz2018IAUS}.
Often the detection rate is on the order of 1\% \citep[e.g.,][]{Sato2005,Bennert2009,Braatz2018IAUS}. 
The \water\ megamaser detection rate toward Compton-thick AGN is much greater $\sim50\%$ \citep{Castangia2019}.
While the sample size is small, the \methanimine\ detection rate is similar toward CONs, $\sim86\%$.
The detection rates of megamasers toward CONs suggests that megamasers may be an indicator of Compton-thick environments; however, a comparison of \methanimine\ emission in a sample of non-Compton-thick galaxies is needed to correctly test this.

The link between the growth of supermassive black holes and  megamasers is well established \citep[e.g.,][]{Reid2009}. 
22~GHz \water\ maser structure has been observed in the accretion disks, jets, and outflows of AGN \citep[e.g.,][]{Henkel2005}. 
Compact OH megamasers trace molecular tori around AGN \citep[e.g.,][]{Lonsdale1998}.
The spectrum of the \methanimine\ megamaser does not yet show extreme velocity components ($>1000$~\kms) similar to the disk 22~GHz \water\ megamasers. 
The \methanimine\ megamasers seem to have more in common with masers observed toward starbursts, jets, or outflows \citep[e.g.,][]{Peck2003,Kondratko2005,konig2017}. 
The \methanimine\ line width in all cases  is $>100$~\kms\ except for IC~860.
The unresolved \methanimine\ megamaser in Arp~220 implies the source is $<103$~pc in diameter (0\farcs32 adopting a distance of 81.8~Mpc) and in IC~860 the source is $<100$~pc (0\farcs35 adopting a distance of 59.1~Mpc).
toward both the eastern and western nuclei of Arp~220, which have respective velocities of 5454~\kms\ and 5342~\kms\ \citep{Martin2016}, the \methanimine\ line is blue shifted by $\sim100$~\kms.
However, these properties are not unique to AGN, unlike extreme velocity components,  and may also result from nuclear starbursts \citep[e.g.,][]{Brunthaler2009,konig2017,Gorski2019}

\subsubsection{36~GHz \methanol\ masers}
\citet{Suzuki2016} hypothesize that \methanimine\ may be abundant toward class I \methanol\ maser sources because both molecules can be formed by hydrogenation of molecules on grain surfaces.
\methanimine\ can be formed by hydrogenation of HCN \citep{Theule2011} and \methanol\  by hydrogenation of CO.
\citet{Suzuki2016} target 12 high-mass star forming regions and two low-mass star forming regions known to have Class I \methanol\ masers. 
\methanol\ masers are divided into two subclasses depending on their pumping scheme.
Class I \methanol\ masers are pumped via collisions and class II \methanol\ masers are radiatively pumped \citep{Menten1991a}. 
\methanimine\ was detected toward eight of the high-mass star forming regions with a fractional abundance in the range of $\sim10^{-9}$ to $\sim10^{-8}$.
The existence of hyper-compact \hii\ regions and weak H54$\beta$, or lack thereof, points to an early stage of high-mass star formation.
Sources with evolved \hii\ regions show less abundance of \methanimine. 
It is suggested from the results of that large \methanimine\ abundances are related to high-mass star formation, and the presence of class I \methanol\ masers.  

\begin{figure}[]
    \centering
    \includegraphics[width=0.48\textwidth]{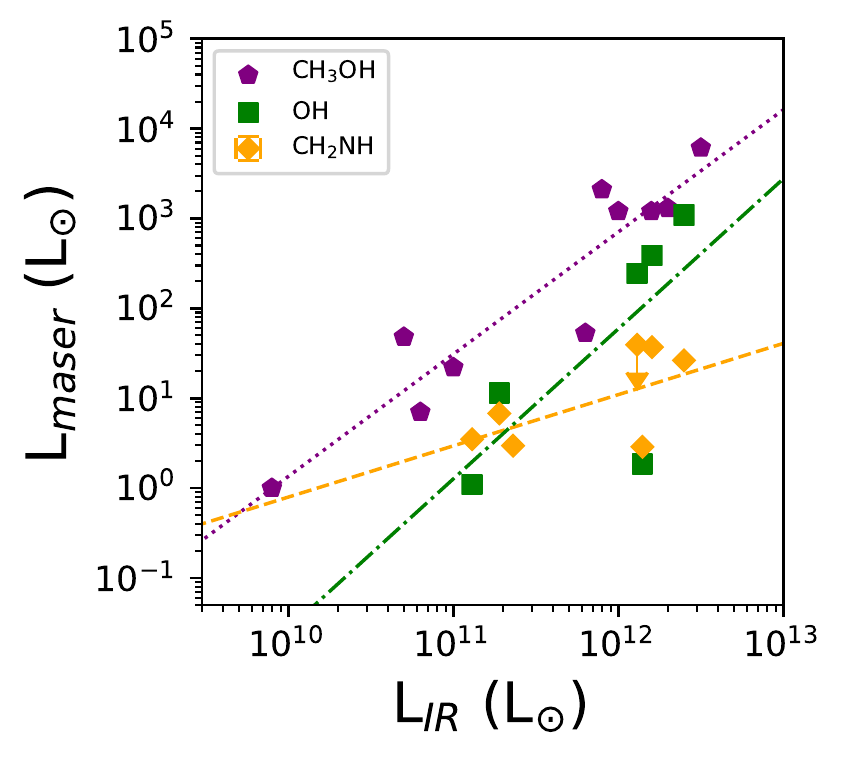}
    \caption{Relationship between the infrared luminosity of the host galaxy and \methanol, OH, or \methanimine\ megamasers.
    Galaxies with 36~GHz \methanol\ masers from \citet{Chen2016} are plotted with purple pentagons, and galaxies with \methanimine\ masers are plotted with orange diamonds.
    OH megamasers for six of the seven CONs (all except IRAS 17578$-$0400) from \citet{Darling2002} and \citet{Wiggins2016} are plotted with green squares.
    The infrared luminosity  of the galaxies are adopted from \citet[][]{Sanders2003}.
    The purple dotted line represents the linear best fit to the \methanol\ data from \citep{Chen2016}, the orange dashed line represents the best fit  to the \methanimine\ data (this paper), and the green dashed dotted line is the linear best fit to the OH maser luminosities from \citet{Darling2002} and \citet{Wiggins2016}.
    }
    \label{fig:methanolmethanimine}
\end{figure}

Since \methanimine\ abundance and class I \methanol\ masers are likely related, we compare extragalactic \methanol\ masers with \methanimine\ masers. 
Figure \ref{fig:methanolmethanimine} shows the infrared luminosity, L$_{\rm IR}$, \citep[][]{Sanders2003} plotted against luminosities of the 36~GHz \methanol\ maser \citep{Chen2016} and 5.29~GHz \methanimine\ maser (this paper) for all presently known extragalactic sources. 
No galaxies have yet been observed in both lines besides Arp~220. 
\citet{Chen2016} note a strong correlation of the 36~GHz \methanol\ maser and infrared luminosity ${\rm L_{methanol}}\propto 1.36\,{\rm L_{IR}}$  (R=0.92):
The strong correlation with the infrared luminosity of the galaxy suggests that the 36~GHz \methanol\ maser is related to star formation processes.
However, when corrected for Malmquist bias the  ${\rm L_{methanol}}$---${\rm L_{IR}}$ relation is shallower with a slope of $1.01\pm0.18$. 
Indeed, when the 36~GHz maser is observed with sufficient angular resolution to be spatially resolved toward star forming galaxies, the maser reveals large-scale shocks \citep[$>10$~pc; e.g.,][]{Gorski2017,Gorski2018,Gorski2019} potentially indicating cloud-cloud collisions. 
If the 5.29~GHz \methanimine\ is linked to class I \methanol\ masers then a similar tight relationship with the infrared luminosity of the galaxy may be observed. 
We find a weak correlation (Fig. \ref{fig:methanolmethanimine}; R=0.78) with the infrared luminosity of the galaxy:
\begin{equation}
     \rm{log}\,L_{\rm{CH_2NH}}[L_\odot]=(0.61\pm0.34)\,\rm{log}\,L_{IR}[L_\odot] - (6.32\pm4.01)
 .\end{equation}
The \methanimine\ maser appears less strongly correlated with the infrared luminosity of the host galaxy.
This may be a result of having a small sample of galaxies or that the \methanimine\ maser traces a different physical process than 36~GHz \methanol\ masers.

\subsubsection{OH masers}
The \methanimine\ maser is emitted from a region smaller than 103~pc and 100~pc respectively toward Arp~220 and IC~860.
OH megamasers are also known to trace the dense molecular environment around CONs.
 \citet{Momjian2006}, \citet{Pihlstrom2001} and \citet{Lonsdale1998} show, with Very Long Baseline Interferometry (VLBI) observations, complicated structures in Arp220, IRAS 17208$-$0014, and III~Zw~35 traced by OH masers. 
The OH  masers trace material near the sphere of influence of the supermassive black hole (e.g., $<30$~pc \citealp{Onishi2017}).
A strong correlation between OH and \methanimine\ masers may indicate the \methanimine\ maser is tracing the feedback resulting from within the CON. 

\begin{figure}[]
    \centering
    \includegraphics[width=0.45\textwidth]{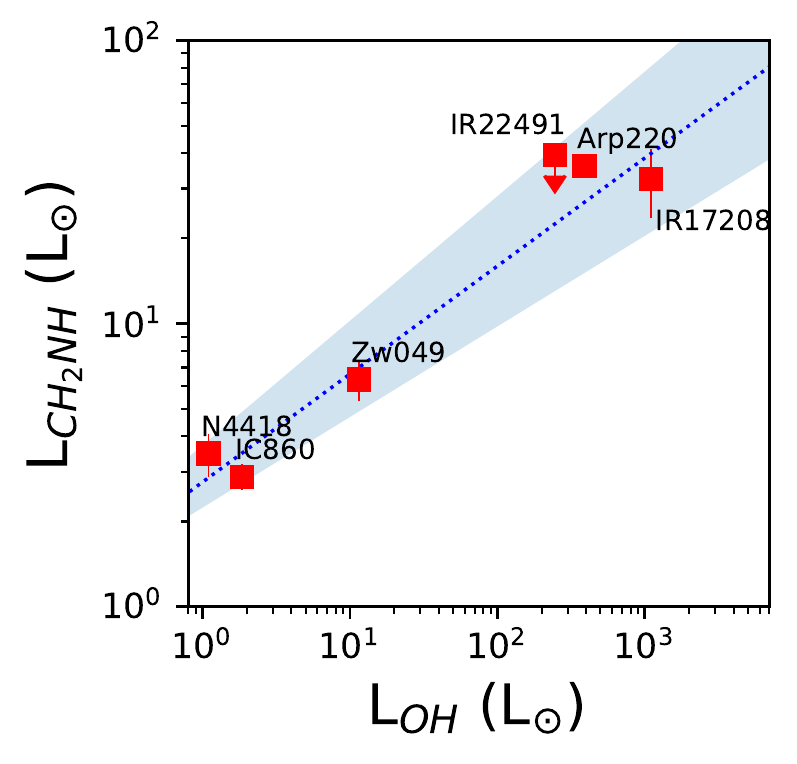}
    \caption{Comparison between the luminosity of the 5.29~GHz \methanimine\ megamasers (this work) and the OH megamasers from \citet{Darling2002} and \citet{Wiggins2016}.
    No OH megamaser has been observed yet in IRAS~17578$-$0400. 
    All galaxies are labeled, and the best fit linear relationship  ($\rm{L}_{\rm{CH_2NH}} \propto \rm{L}_{\rm{OH}}^{0.36\pm0.05} $, R=0.97) is shown with a blue dotted line. The shaded blue area represents the uncertainty in the best fit. 
    }
    \label{fig:OH-methanimine}
\end{figure}

For the CONs we find that the OH maser luminosity scales with infrared luminosity by (Fig. \ref{fig:methanolmethanimine}; R=0.84):
\begin{equation}
     \rm{log}\,L_{\rm{OH}}[L_\odot]=(1.66\pm0.84)\,\rm{log}\,L_{IR}[L_\odot] - (18.2\pm10.0)
 .\end{equation}
The luminosities for our sample of CONs  of OH megamasers are adopted from \citet{Wiggins2016} except for  IRAS 22491$-$1808, which is adopted from \citet{Darling2002}, and IRAS~17578$-$0400, for which an OH detection is absent in the literature.
This is consistent with the relationship found by surveys of IRAS galaxies by \citet{Kandalian1996,Darling2002}. 

We find a strong correlation between the 5.29 GHz \methanimine\  and OH maser luminosities (Fig. \ref{fig:OH-methanimine}; R=0.97),
\begin{equation}
     \rm{log}\,L_{\rm{CH_2NH}}[L_\odot]=(0.36\pm0.05)\,\rm{log}\,L_{OH}[L_\odot] - (0.45\pm0.09)
 ,\end{equation}

toward Arp~220 \citep{Lonsdale1998}  compact  ($<$1~pc) and diffuse OH maser emission is revealed.
The compact OH maser regions appear to be pumped by collisions, whereas the diffuse masers are pumped via infrared radiation, though \citet{Pihlstrom2001} and \citet{Parra2005} argue  that the observed differences in the line ratios between compact and diffuse phases could be a natural property of one phase characterized by clumpy unsaturated masers.
The strong correlation between OH and \methanimine\ maser luminosities, and the spatial coincidence within  $\sim100$~parsecs,  suggests they trace similar processes in the CONs.
The critical density of the upper 1$_{10}$ state of \methanimine\ is estimated as $\sim1\times10^5$~cm$^{-3}$ \citep{Faure2018}, whereas OH mases at densities an order of magnitude lower $>10^4$~cm$^{-3}$ \citep{Baan1991}.
Consequently, the \methanimine\ maser may trace denser structures in CONs.
Parsec scale resolution observations of the \methanimine\ maser may provide insights to the nature of these massive structures.

\subsection{Maser luminosity and infrared relationship}

We found that the \methanimine\ maser is weakly correlated with the infrared luminosity of the host galaxy.
Usually, the luminosity of a radiatively pumped maser is proportional to the availability of pumping photons and stimulating photons, for example $L_{\rm{maser}} \propto L_{\rm{stim}}L_{\rm{pump}}$ \citep{Baan1989}.
However, one can, perhaps naively, assume that for a low-gain maser pumped by the infrared radiation field,  that is stimulated by the radio continuum and unsaturated,  the luminosity of the maser is proportional to the square of the infrared luminosity (e.g., $L_{\rm{maser}} \propto L_{\rm{IR}}^2$). 
This is because the stimulating photons from the radio continuum are proportional to the infrared flux.
Saturated masers no longer grow in luminosity exponentially, so the luminosity of the maser will be proportional to the infrared luminosity (e.g., $L_{\rm{maser}} \propto L_{\rm{IR}}$).
Thus, for an unspecified number of radiatively pumped masers, we expect the luminosity of the maser to be proportional to  $L_{\rm{IR}}^\alpha$ where $\alpha$ has a value  between 1 and 2 (see Sect. 5.4 of \citealp{Darling2002} and references therein for a more detailed discussion).
We observe that the \methanimine\ masers are under luminous for infrared pumping in this scenario with $L_{\rm{CH_2NH}} \propto {0.61\pm0.34}\, L_{\rm{IR}}$.
We can imagine a few scenarios that might explain this situation. 

First, the maser may still be radiatively pumped, but at millimeter-to-submillimeter wavelengths.
A lack of infrared pumping is plausible, as much of the mid-infrared is attenuated and reemitted at longer wavelengths \citep{Aalto2015a,Aalto2019}.
Thus giving rise to a weaker correlation between the maser and the infrared luminosity of the galaxy.
Our models show that pumping mainly occurs from the intense radiation at millimeter-to-submillimeter wavelengths, supporting this scenario.

Second, the \methanimine\ maser is not subject to the total infrared radiation field, but still radiatively pumped. 
Perhaps the regions responsible for the \methanimine\ maser are shielded from the intense radiation from the CON environment, and they are pumped by a more local source such as a nearby star-forming region.
As these regions also excite OH masers this gives rise to the strong correlation between \methanimine\ and OH masers. 
In this scenario \methanimine\ absorption lines would likely be observable in the mid-infrared.

Last, the \methanimine\ molecules do not experience the radiation field and the maser is collisionally pumped.
The radiation field is either too heavily shielded or too dilute to pump the molecules.
The \methanimine\ maser is then related to the same physical process as the collisionally pumped OH masers yielding a strong correlation.
\citet{Lonsdale1998} suggesting that shock fronts from molecular tori around newly formed AGN could result in these spectrally broad features.

\section{Conclusions}
We have conducted the first search for the 5.29 GHz \methanimine~$1_{10}-1_{11}$ transition in a sample of galaxy nuclei using the VLA. 
\methanimine\ emission is detected toward six out of seven galaxies with CONs: Zw~049.057, IRAS~17208$-$0014 IRAS~17578$-$0400, NGC~4418, IC~860, and Arp~220. 
\form\ emission is also detected toward three galaxies: Zw~049.057,  IRAS~17578$-$0400, and IC~860. 
In all observations, the emission is spatially unresolved.

The \methanimine\ emission detected toward the western core of Arp~220 has an isotropic luminosity of 27~\Lsun\ and a brightness temperature $> 1400$~K, providing evidence for the first \methanimine\ megamasers. 
The isotropic luminosities measured toward the other galaxies range from 2$\sim$10~\Lsun; however, the spatial resolution of the observations only allows for lower limits of the brightness temperature.
Non-LTE modeling suggests that the \methanimine\ maser is pumped by the intense millimeter-to-submillimeter radiation field, though pumping from collisions cannot be excluded.
Currently, the structure is measured to be smaller than 103~pc toward Arp~220 and smaller than 100~pc toward IC~860, which is consistent with  the HCN-VIB emitting regions of CONs ($\lesssim$60~pc \citealp{Aalto2015b,Aalto2019}).

Our investigation reveals that the \methanimine\ masers are weakly correlated with the infrared luminosity of the galaxy and strongly correlated with OH megamaser luminosities. 
We hypothesize that the strong correlation between \methanimine\ masers and OH masers is due to collisions in the molecular tori around embedded AGN, although other explanations are possible. 
In this picture, \methanimine\ is shielded from the intense infrared radiation field in CONs, giving rise to a weak correlation between the \methanimine\ maser luminosity and the global infrared luminosity of the host galaxy.
Higher angular resolution observations are needed to reveal the physical structure of the emission and to reveal the pumping mechanism.
\highlight[comment=~]{Altogether}, \methanimine\ megamasers  provide  a  new  tool  for investigating the nuclear processes in galaxies and a potential avenue for observing parsec-scale structure in CONs.  
 
\begin{acknowledgements}

S.A., M.G., K.O., S.K. N.F. gratefully acknowledges support from an ERC Advanced Grant 789410 a.

The National Radio Astronomy Observatory is a facility of the U.S. National Science Foundation operated under cooperative agreement by Associated Universities, Inc.
\end{acknowledgements}

\bibliographystyle{aa_url} 
\bibliography{methanimine,maser,Con Quest,misc,jb-add} 

\begin{appendix}
\section{ Molecular data for CH$_2$NH and H$_2$CO}
\begin{figure}
   \centering
    \includegraphics[width=0.495\textwidth]{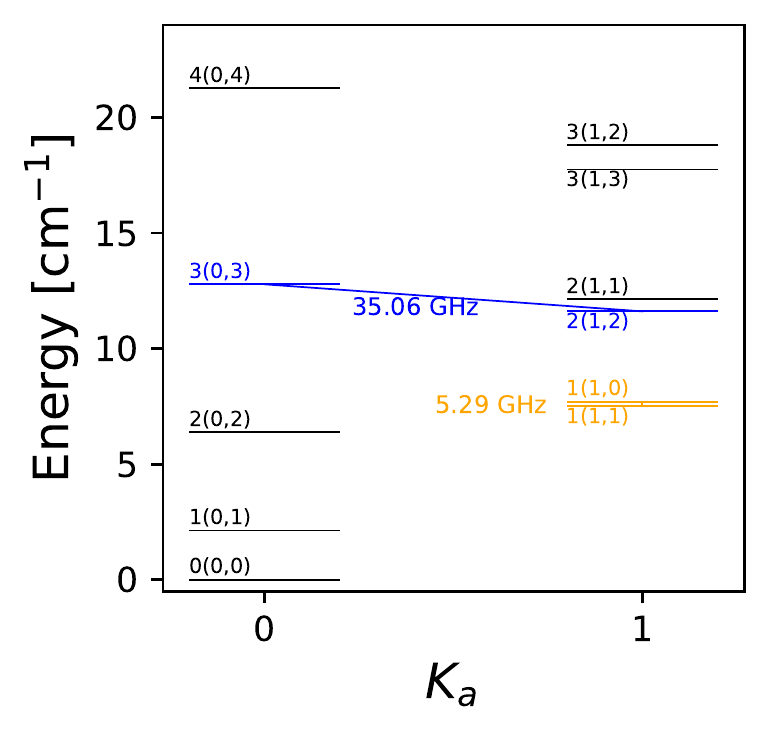}
    \vspace{-0.2in}
    \caption{ Energy level diagram of \methanimine\ showing the first 11 energy levels.
    Each state is labeled with quantum numbers $J(K_a,K_c).$
    The 5.29~GHz line is identified in orange, and the 35.05~GHz line is identified in blue.
    }
    \label{fig:energy-levels}
    \end{figure}

Methanimine provides a useful test case of the effects of radiative excitation of an interstellar molecule that is known to exhibit maser emission. 
The molecule has a relatively large permanent dipole moment in its ground state, $\sim 2$ Debye, with comparable components along the a- and b- principal axes of rotation. 
Thus, it has relatively strong rotational transitions at both centimeter and millimeter wavelengths.  
Figure \ref{fig:energy-levels} shows the lowest 11 energy levels with the predicted maser transitions at 5.29~GHz and 35.05~GHz respectively labeled in orange and blue. 
All nine of the vibrational modes are infrared-active, and all of the vibrational fundamental bands have been rotationally analyzed. 
The rotational analyses of the infrared spectra of \methanol\ were carried out more than 30 years ago, mainly by Duxbury and Halonen and their collaborators through the use of the high-resolution  Fourier-transform spectrometer at the McMath Solar Telescope on Kitt Peak (see Table \ref{tab:specdat}).
We have used a slightly modified version of {\tt asymbd}, a computer code originally written by Arthur Maki in the 1970s, to compute the rotational and vibration-rotation spectra of \methanimine\ (A.~G.~Maki, 1986, private communication to R. Bumgarner and J. H. Black). 
All of these computations have used the A reduction of the Hamiltonian operator in the 1R representation. 
The inputs are the rotational constants and band-head frequencies taken from the spectroscopic analyses. 
The spontaneous transition probabilities of pure rotational transitions are based on the accurate values of the ground-state permanent dipole moments,
$\mu_a=1.3396$ and $\mu_b=1.4461$~D   \citep{1979JChPh..70.2829A}.  

For the vibration-rotation transition probabilities, we adopted the band strengths of \cite{2016PCCP...18.4201C}, 
which were determined from {\it ab initio} calculations. 
They compared band strengths obtained with several different treatments of the geometric  derivative tensors of the electric dipole moment. We adopt the results of their numeric, anharmonic method with a step size of 0.01 \AA. 
Cornaton et al. did not tabulate their computed band strengths or transition moments, but rather presented a graph of their calculated infrared spectra of \methanimine. In the calculated spectrum, each band was represented by a Lorentzian profile of FWHM $\Gamma = 10$ cm$^{-1}$. Although their graph of the spectrum has a vertical axis labeled ``Intensity / km mol$^{-1}$,'' it can be deduced that this must be the ``intensity'' per unit wavenumber and that the integrated band intensity $S_{\rm band}$ [km/mole] should be given by 
$$  S_{\rm band} = \pi \Gamma I_{\rm peak} / 2, $$
where $I_{\rm peak}$ is the peak intensity displayed for each band. It is also important to note that Cornaton et al. labeled some of the bands differently from the spectroscopists: The vibrational modes $\nu_1$ through $\nu_6$ are in the same order. 
The spectroscopic analyses distinguished between the first seven modes of A$'$ symmetry (mixed a- and b-type spectra for in-plane vibrations) and the last 2 modes of A$''$ symmetry (c-type spectra for out-of-plane vibrations); thus modes $\nu_7$, $\nu_8$, and $\nu_9$ were listed by Cornaton et al. as $\nu_9$, 
$\nu_7$, and $\nu_8$, respectively. 
We adopt the ordering from the spectroscopic literature. 
The dimensionless absorption oscillator strength $f$ of a band is directly related to the integrated intensity,
$$ S_{\rm band} = 5.33129\times 10^6 f \;\;\; {\rm [km/mole],} $$
and the dipole transition moment in Debye can be written as \comment[]{fixed $\tilde{\nu}$ not rendering}
$$  \mu(v',v'') = \bigl( f / 4.70176\times 10^{-7} \tilde{\nu}  \bigr)^{1/2} \;\;\;{\rm [D]}  $$
when the degeneracies of the upper ($v'$) and lower ($v''$) states are the same. 
Our adopted values of the transition moments are in good agreement with the older {\it ab initio} calculations of
\cite{1984JMoSp.105...70H}
and with the relative intensities determined from measurements of 
$\nu_7$, $\nu_8$, and $\nu_9$ by
\cite{1986JChPh..85..692H}, and 
\cite{1986ApJ...303..897H}.
Table {A.1} contains our adopted
values for the nine fundamental vibration-rotation bands. The energy $E_i$ of the $J(K_a,K_c) = 0(0,0)$ 
level of each $v_i=1$ state is listed, together with the inverse lifetime at low $J$, $1/t(v_i,u) = 
\sum_{\ell} A(v_i=1,u;v=0,\ell)$, where $A$ is the spontaneous transition probability in [s$^{-1}$] for
a $v_i=1 \to v=0$ band and $u$ and $\ell$ stand for the upper and lower rotational quantum numbers,
respectively. The final column lists references to the sources of spectroscopic constants.
%
\begin{table*}
\caption{Adopted spectroscopic data and intensities for \methanimine.}
\begin{center}
\begin{tabular}{clrrlll}
\hline
 & Mode (i) & $E_i(0_{0,0})$ & $A=1/t_i$ & $I_{\rm peak}$ & $\mu(i)$ & References \\
 &  & cm$^{-1}$ & s$^{-1}$ & km/mole/cm$^{-1}$ & D & \\
 \hline
 & $v=0$ & 0.000 & \dots & \dots & 1.971 & 1,2 \\
 & $\nu_1$  & 3262.622 & 4.24 & 0.202 & 0.0198 & 3 \\
 & $\nu_2$ & 3024.452 & 50.86 & 2.764 & 0.0769 & 4 \\
 & $\nu_3$ & 2914.184 & 60.57 & 3.551 & 0.0884 & 4 \\
 & $\nu_4$ & 1638.299 & 7.41 & 1.448 & 0.0735 & 2,5 \\
 & $\nu_5$ & 1452.048 & 1.71 & 0.419 & 0.0423 & 5,6 \\
 & $\nu_6$ & 1344.261 & 11.72 & 3.257 & 0.1237 & 5,6 \\
 & $\nu_7$ & 1058.181 & 5.60 & 2.495 & 0.1213 & 1,7,8,9 \\
 & $\nu_8$ & 1126.988 & 7.06 & 2.914 & 0.1267 & 1,7,8,9  \\
 & $\nu_9$ & 1060.760 & 2.47 & 1.162 & 0.0821 & 1,7,8,9  \\
\hline
\end{tabular}
\label{tab:specdat}
\end{center}
\tablefoot{References: 
[1]  \cite{1985JChPh..83.2078H} ; 
[2] \cite{1979JChPh..70.2829A}; 
[3] \cite{1985CPL...118..246H}; 
[4] \cite{1985JChPh..83.2091H}; 
[5] \cite{DC9817100097}; 
[6] \cite{1982JMoSp..92..326D}; 
[7] \cite{1986ApJ...303..897H}; 
[8] \cite{1986JChPh..85..692H}; 
[9] \cite{1984InfPh..24..261D}.
}
\end{table*}

A line list with transition probabilities has been computed for each of the pure-rotational spectra 
(ground state and nine $v=1$ states) and for each of the nine fundamental bands, that is, 19 lists in all.
The maximal rotational quantum number is $J_{\rm max} = 29$ and the ground-state rotational energies are 
all less than 1000 cm$^{-1}$, so that the $v=0$ levels will not overlap with excited-state levels. The 
19 lists were combined to form a single file in the format of the Leiden Atomic and Molecular Database
({\tt LAMDA}; see \citealp{2005A&A...432..369S})\footnote{{\tt https://home.strw.leidenuniv.nl/$\sim$moldata/}}, 
with an energy-ordered list of term energies, a frequency-ordered table of transitions, and tables of 
downward inelastic collisional rate coefficients. The combined list contains only excited levels
that can be reached by radiatively allowed transitions from $v=0$ levels. 
The computing time required to
solve the non-LTE rate equations increases rapidly with the number of levels. Therefore, several versions 
of the {\tt LAMDA}-format file were constructed, as summarized in Table \ref{tab:moldat}.
\begin{table*}

\caption{ Properties of the molecular data files for \methanimine.}
\begin{center}
\begin{tabular}{crrrrl}
\hline
 & Levels & Transitions & $J_{\rm max}$ & highest level & $E_{\rm max}$ \\  
 &            &                    &                         & $v=0,J_{K_a,K_c}$        & [cm$^{-1}$]  \\
 \hline
 vibrot &  all $v$ & && &   \\
 & 360 & 3326 & 5 & 5(5,0)  & 168.813 \\  
 & 1210 & 16835 & 10 & 10(10,1) & 662.827 \\  
 & 2542 & 40852 & 15 & 13(12,1) & 978.137 \\  
 & 4126 & 69226 & 20 & 20(10,10)  & 993.167 \\  
 & 6633 & 111780 & 29 & 20(10,10) & 993.167 \\  
 hfs &  $v=0$ only & & & & \\
 &1446 & 3209 & 29 & 29(9,21) & 1369.959 \\  
\hline
\end{tabular}
\end{center}
\tablefoot{ \noindent The available data files for CH$_2$NH are characterized by (Col. 1) the number of vibrational states
included, (2) the number of levels, (3) the number of radiative transitions, (4) the largest value of the rotational quantum number $J_{\rm max}$ , (5) the highest level in the vibrational
ground state,  and (6) the energy of the highest level in the vibrational ground state.}
\label{tab:moldat}
\end{table*}
The smallest data file is not recommended for general use, although it might be valuable for tracing
the parameter space that allows strong maser action, where excitation computations converge
very slowly. 

Hyperfine structure (HFS) in low-$J$ transitions of \methanimine\ is well resolved in laboratory spectra and in the spectra of narrow-line astronomical sources. 
In low-lying $\Delta J = 0$ transitions, the HFS is quite large; for example, six HFS components are spread over 154 km s$^{-1}$ in Doppler velocity around the $1_{10} - 1_{11}$ transition at 5.290 GHz. 
In the higher frequency transitions at 200 to 400 GHz, the HFS covers only 1 to 3 km s$^{-1}$. 
We neglect HFS entirely in the data files that include vibrationally excited states. 
The last line of Table \ref{tab:moldat} refers to a data file with HFS in transitions within the ground vibrational state only. 
This file has been constructed directly from the HFS-resolved file {\tt c029518{\_}hfs.cat} in the Cologne Database for Molecular Spectroscopy (CDMS; see M{\"u}ller et al. 2001).\footnote{CDMS at {\tt https://cdms.astro.uni-koeln.de/classic/entries/}}

\citet{Faure2018} employed high-level quantum mechanical methods to compute
the interaction potential and collision dynamics of para-H$_2$ colliding with \methanimine. Their work gave cross sections and rate coefficients for inelastic collisions involving the 15 lowest rotational levels of \methanimine\  with energies $E < 28.3$ cm$^{-1}$ (or $E/k < 41$ K). 
They included these rates in a non-LTE model of the excitation of \methanimine\  to predict weak maser emission in the 5.29 GHz transition, with an intensity in good agreement with that observed in the molecular cloud Sgr B2(N) near the Galactic Center. 
The full set of computed collision rates was not published, although rates for four collisional transitions were displayed in Fig. 2 of \citet{Faure2018}. 
The rate coefficient for the downward transition between the two lowest rotational levels, $u\to \ell = 1_{01} \to 0_{00}$, is $q_{u\ell} = 5.5\times 10^{-11}$ cm$^3$ s$^{-1}$ at a kinetic temperature $T=30$ K. 
Without a set of accurate collision rates, we cannot explore effects of non-LTE excitation that are sensitive to details of the state-to-state collision-induced transitions. 
However, we can use an extensive set of very crude collision rates to investigate to order-of-magnitude the competition between collisional and radiative processes. 
All of the data files listed in Table 2  contain collision rates calculated with the following simple prescription for radiatively allowed transitions only. 
The downward collision rate coefficient $q_{u\ell}$ for a transition $u \to \ell$ is given by 
$$  q_{u\ell} = q_0 S_{u\ell} / \sum_i S_{ui}, $$
where $S_{u\ell}$ is the dimensionless radiative line strength computed in the program {\tt asymbd} as an intermediate step toward the spontaneous transition probability. 
We adopt $q_0 = 5.5\times 10^{-11}$~cm$^3$~s$^{-1}$ for pure rotational transitions within a given vibrational state, which recovers the rate computed by \citet{Faure2018} for the $1_{01} \to 0_{00}$ transition for which $S=1.0$. 
For all transitions in which the vibrational quantum number changes, we adopt a scaling factor that is an order of magnitude smaller, $q_0 =  5.5\times 10^{-12}$ cm$^3$ s$^{-1}$.  
These adopted downward rates are taken to be independent of temperature. 
The upward rates are computed from the downward rates in detailed balance within the {\tt GROSBETA} program. The collision partner is a generic neutral species, namely the hydrogen molecule. 
The adopted collision rates neglect  $\vert \Delta J \vert > 1$ transitions, which are expected to be important in such an asymmetric, polar molecule as \methanimine. 
A detailed summary of recent progress in the study of collisional processes is given in the review of the {\tt LAMDA} database by \citet{vanderTak2020}. 
The crude collision rates adopted here limit the accuracy of any deduced hydrogen density to an order of magnitude. 
As suggested in the main text, these rates may still be adequate to draw broad conclusions about the parameter space of excitation in real sources, especially where radiative processes dominate. 

In the case of formaldehyde, H$_2$CO, a similar data file has been constructed based upon the
entry in the CDMS for the ground vibrational
state and on the entry in the HITMAN database \cite{2017JQSRT.203....3G}\footnote{{\tt https://hitran.org/}} for vibrationally excited states. 
This data set comprises 1034 vibrational-rotational levels with maximal rotational quantum number $J_{\rm max}=13$ and energies up to 3304 cm$^{-1}$.
The line list of 3791 transitions spans frequencies from 1 MHz to 93 THz. 
Rate coefficients for inelastic collisions of para-H$_2$ and ortho-H$_2$ with H$_2$CO have been taken from \cite{2013MNRAS.432.2573W}.
These rates apply for temperatures between 10 K and 300 K. 
The ortho and para nuclear-spin species of H$_2$CO are treated together,
their relative abundances being determined by the value of an additional parameter, the formation
temperature, $T_{\rm form}$. Although it is possible to include reactive ortho/para interchange 
processes, these are not used in the present calculations. In all the results discussed here, 
$T_{\rm form} = 300 K$, which is high enough to ensure that the ortho/para ratio approaches the
ratio 3 of the corresponding nuclear-spin statistical weights.

\section{Continuum radiation and pumping rates of CH$_2$NH in IC~860}

\begin{figure}
   \centering
   \includegraphics[width=0.495\textwidth]{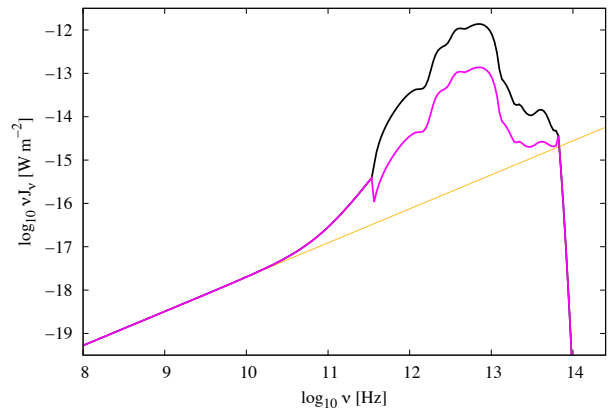}
   \vspace{-0.4in}
    \caption{Internal spectral energy distribution of IC~860 with an effective solid angle $\Omega=3\times 10^{-12}$ sr and a total power $L=6\times 10^{11}$  L$_{\odot}$. 
    The upper curve (black in the submillimeter and infrared $\log\nu=11.5$ to $13.8$) shows the brightness for $\varphi=1$, while the lower curve (magenta in the submillimeter and infrared) is the corresponding brightness for $\varphi=0.1$.
    The orange line indicates the underlying nonthermal power-law source, which dominates at radio frequencies.
    }
    \label{SED_nujnu}
   \end{figure}

\begin{figure}
   \centering
    \includegraphics[width=0.495\textwidth]{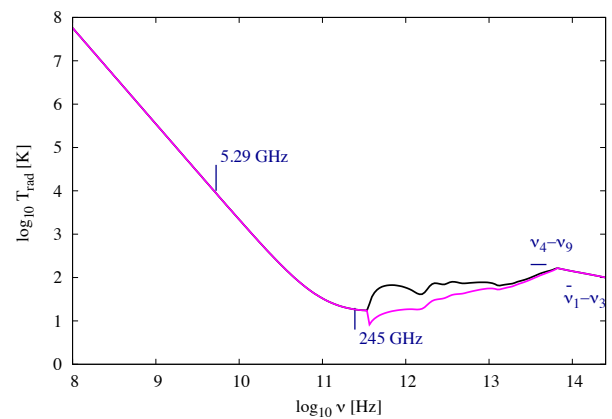}
    \vspace{-0.4in}
    \caption{Corresponding internal  spectral energy distribution of IC~860 presented in the form of a radiation brightness temperature $T_{\rm rad}$. 
    As in Fig. \ref{SED_nujnu}, the brightness in the infrared-to-submillimeter is elevated at $\varphi=1$ (black curve) compared to $\varphi=0.1$ (magenta) throughout the submillimeter-to-infrared region.
    Frequencies of 5.29 GHz (where $T_{\rm rad}\approx 8791$ K) and 245 GHz  (where $T_{\rm rad}\approx 18$ K) are marked. Short horizontal bars indicate the frequencies covered by the $\nu_1$ to $\nu_3$ and the $\nu_4$ to $\nu_9$ vibrational fundamental bands of CH$_2$NH, where $T_{\rm rad}\approx 100$ K.}
    \label{SED_Trad}
    \end{figure}

As described in Sect.~\ref{sec:LVG}, a realistic internal radiation field is an essential element of the non-LTE
radiative transfer models for molecules such as CH$_2$NH and H$_2$CO. We explored 
empirical models of the continuum spectral energy distribution for IC~860 in some detail.
The model radiation field is displayed in Figs. \ref{SED_nujnu} and \ref{SED_Trad} in different forms for two values of $\varphi = 1$ and $0.1$. 
Figure \ref{SED_nujnu} shows $\log \nu J_{\nu}$ versus $\log \nu$. 
Figures \ref{SED_Trad} displays the equivalent radiation brightness temperature $T_{\rm rad}$. 
We note that the nonthermal power-law component dominates at radio frequencies.

It is useful to compare the spontaneous and stimulated emission rates in the model
radiation field for selected transitions of CH$_2$NH over a wide range of frequencies.
These rates are collected in Table \ref{tab:pumping} to illustrate the competition between processes
that deplete or add to the population of the $1_{10}$ and $1_{11}$ levels. We note in
particular that the 5.29 GHz transition itself, $u=1_{10}\to \ell=1_{11}$, suffers absorption
and stimulated emission in the adopted radiation field at rates that are $3.5\times 10^4$
times faster than the spontaneous transition probability. The total spontaneous decay
rates out of the $1_{10}$ and $1_{11}$ levels are $5.65\times 10^{-5}$ s$^{-1}$ and
$9.33\times 10^{-5}$ s$^{-1}$, respectively. Thus, population is drained from the lower
of the two levels faster than from the higher one. As a result, the populations of these levels will
be automatically inverted (negative excitation temperature) if collisions are relatively
unimportant and if the rates are comparable for the processes that populate them. 
It is also apparent that the most rapid radiative processes in this radiation field tend to
be the rotational transitions at millimeter and submillimeter wavelengths. Further observations of
the millimeter-submillimeter-wave continuum with high angular resolution will make it possible to
refine the description of the radiation field.

\begin{table*}
 \caption{ Selected pumping rates of CH$_2$NH in the IC~860 radiation model.}
  \begin{center}
   \begin{tabular}{lccrlrlrl}
   \hline\\
      &     &        &       &              & $\varphi=1.0$ && $\varphi=0.1$ \\
 Band & $u$ & $\ell$ & $\nu$ & $A_{u,\ell}$ & $T_{\rm rad}$ & $y(\nu) A_{u,\ell}$ & $T_{\rm rad}$ & $y(\nu) A_{u,\ell}$ \\
      &     &        & [GHz] & [s$^{-1}$]   &   [K]         &  [s$^{-1}$]   &   [K]         &  [s$^{-1}$] \\
   \hline\\
$v=0$   & $1_{10}$ & $1_{11}$ &     5.29 & $1.55\times 10^{-9}$ & 8791.0 & $5.35\times 10^{-5}$ & 8791.0 & $5.35\times 10^{-5}$ \\
$v=0$   & $2_{11}$ & $1_{10}$ &   133.27 & $1.48\times 10^{-5}$ &   26.1 & $5.33\times 10^{-5}$ &   26.1 & $5.33\times 10^{-5}$ \\
$v=0$   & $1_{10}$ & $1_{01}$ &   166.85 & $5.64\times 10^{-5}$ &   22.2 & $1.30\times 10^{-4}$ &   22.2 & $1.30\times 10^{-4}$ \\
$v=0$   & $2_{21}$ & $1_{10}$ &   617.87 & $1.72\times 10^{-3}$ &   66.5 & $3.07\times 10^{-3}$ &   16.0 & $4.59\times 10^{-4}$ \\
$\nu_9$ & $0_{00}$ & $1_{10}$ & 31569.95 & $2.47$ &   99.6 & $6.09\times 10^{-7}$ &   89.0 & $9.96\times 10^{-8}$ \\
$\nu_7$ & $1_{11}$ & $1_{10}$ & 31726.82 & $2.81$ &  100.0 & $6.83\times 10^{-7}$ &   90.2 & $1.30\times 10^{-7}$ \\
$\nu_8$ & $0_{00}$ & $1_{10}$ & 33555.42 & $7.06$ &  104.7 & $1.48\times 10^{-6}$ &   94.4 & $2.76\times 10^{-7}$ \\
$\nu_6$ & $1_{11}$ & $1_{10}$ & 40301.39 & $5.83$ &  121.7 & $7.30\times 10^{-7}$ &  109.9 & $1.33\times 10^{-7}$ \\
$\nu_5$ & $1_{11}$ & $1_{10}$ & 43524.31 & $8.60\times 10^{-1}$ &  128.4 & $7.40\times 10^{-8}$ &  116.6 & $1.43\times 10^{-8}$ \\
$\nu_4$ & $1_{11}$ & $1_{10}$ & 49109.62 & $3.72$ &  137.8 & $1.39\times 10^{-7}$ &  127.3 & $3.39\times 10^{-8}$ \\
$\nu_3$ & $1_{11}$ & $1_{10}$ & 87360.56 & $30.4$ &  147.3 & $1.32\times 10^{-11}$ & 147.3 & $1.32\times 10^{-11}$ \\
$\nu_2$ & $1_{01}$ & $1_{10}$ & 90503.71 & $25.5$ &  145.4 & $2.71\times 10^{-12}$ & 145.4 & $2.71\times 10^{-12}$ \\
$\nu_1$ & $1_{11}$ & $1_{10}$ & 97805.61 & $2.12$ &  141.3 & $7.95\times 10^{-15}$ &  141.3 & $7.95\times 10^{-15}$  \\

    \hline\\
   \end{tabular}
  \end{center}
  \tablefoot{In the first column, Band identifies either purely
rotational transitions within the vibrational ground state ($v=0$) or the fundamental band of vibrational mode $\nu_i$. The rotational quantum numbers of the upper and lower states of each transition are labeled $u$ and $\ell$, respectively. The frequency of a transition is $\nu$ and the corresponding spontaneous transition probability is $A_{u,\ell}$. Radiation brightness temperatures, $T_{\rm rad}$, and rates of stimulated emission, $y(\nu)A_{u,\ell}$, are tabulated for two versions of the model of the continuous radiation in the center of IC 860. These radiation fields are distinguished by the value of $\varphi$, which describes the fraction of the observed infrared power that is contained within the solid angle subtended by the centimeter-wave radio source in the core.}
  \label{tab:pumping}
\end{table*}
\end{appendix}

\end{document}